\documentclass[10pt]{article}
\begin{document}
%%%%%%%%%%%%%%%%%%%%
\title{{\bf Hawking radiation in GHS blackhole, Effective action and Covariant 
Boundary condition}}
%%%%%%%%%%%%%%%%%%%%
\author{
{Sunandan Gangopadhyay$^{}$\thanks{sunandan@bose.res.in}}\\ 
S.~N.~Bose National Centre for Basic Sciences,\\JD Block, 
Sector III, Salt Lake, Kolkata-700098, India\\[0.3cm]
}
\date{}

\maketitle

%%%%%%%%%%%%%%%%
\begin{abstract}
%%%%%%%%%%%%%%%%
\noindent We exploit the expression for the anomalous (chiral) effective
action to obtain the Hawking radiation from the GHS (stringy) blackhole 
falling in the class of the most general spherically symmetric blackholes
$(\sqrt{-g}\neq1)$, using only covariant boundary condition at the event
horizon. The connection between the anomalous and the normal energy-momentum
tensors is also established from the effective action approach. 
\\[0.3cm]
\\[0.3cm]

\end{abstract}
%%%%%%%%%%%%%%%%%%%%%%%%%%%%%%%%%%%%%%%%%%%%%%%%%%%%%%%%%%%%%%%

\noindent{\it Introduction:}\\

\noindent There are several derivations of Hawking
radiation. The most direct is Hawking's original one
\cite{Hawking1,Hawking2} which computes the Bogoliubov
coefficients between in and out states for a body collapsing
to form a blackhole. Another elegant derivation based
on Euclidean quantum gravity \cite{Hawking3} has been interpreted
as a calculation of tunnelling through classically forbiden
trajectories \cite{parikh}. Therefore it remains of interest to
consider alternative derivations since although there are
many approaches, none is completely clinching or conclusive.

\noindent Recently, a new derivation of the Hawking effect 
has been given in \cite{Robwilczek,Isowilczek}. It relies on
the cancellation of anomalies\footnote{For a detailed discussion
of anomalies see \cite{bert,fujikawa}.} (gravitational and
gauge) at the horizon.
It was shown that effective field theories 
become two dimensional and chiral near the 
event horizon of a black hole by the process of 
dimensional reduction.
This leads to the occurrence of gravitational and gauge anomalies. 
The Hawking flux is necessary to cancell these anomalies.
Further applications of the approach was made in 
\cite{Muratasoda1}-\cite{Isomorita}.

\noindent The approach of \cite{Robwilczek,Isowilczek} was 
further generalised in \cite{shailesh} where it was shown 
that unlike in \cite{Robwilczek,Isowilczek}, 
the complete analysis was feasible 
in terms of covariant expressions only.
The flux from a charged black hole was correctly 
determined by a cancellation of the covariant anomaly 
with the boundary condition being the vanishing of the covariant current 
(and energy-momentum tensor) at the horizon.
The method was soon extended in \cite{sun} 
to discuss Hawking radiation
from the Garfinkle-Horowitz-Strominger (GHS) blackhole spacetime
in string theory which is an example of the 
most general spherically symmetric blackhole ($\sqrt{-g}\neq1$)
\cite{gibb,ghs}. 

\noindent From the analysis of \cite{Robwilczek,Isowilczek,shailesh,sun} 
it appears therefore that covariant boundary conditions 
at the horizon play a fundamental
role. Indeed in \cite{rbshailesh},  
the arguments of \cite{Robwilczek,Isowilczek} which imply that effective
field theories are chiral and two dimensional near the horizon 
was adopted. Then by exploiting known structures of the two dimensional 
effective actions, deductions for the currents 
and the energy-momentum tensors have been made
by the imposition of covariant
boundary conditions only at the horizon. 
The Hawking flux from charged black holes is 
correctly reproduced in this manner. 

%%%%%%%%%%%%%%%%%%%%%%%%%%%%%%%%%%%%%%%%%%%%%%%%%%%%%%%%%%%%

\noindent In this paper, we shall adopt the 
method in (\cite{rbshailesh}) to discuss
Hawking radiation from the GHS blackhole. However, we shall ignore
effects to the Hawking flux due to scatterings by the gravitational
potential, for example the greybody factor \cite{kim}.\\ 
%%%%%%%%%%%%%%%%%%%%%%%%%%%%%%%%%%%%%%%%%%%%%%%%%%%%%%%%%%%%
\newpage
\noindent{\it General Setting and Effective Action :}\\

\noindent We are interested in discussing Hawking 
radiation from GHS blackhole\footnote{The GHS 
blackhole is a member of a family of solutions
to low-energy string theory described by the action (in the string frame)
$S = \int d^{4}x \sqrt{-g} e^{-2\phi}
\left[ -R -4(\nabla\phi)^2 + F^{2}\right]$,
where $\phi$ is the dilaton field and $F_{\mu\nu}$ is the Maxwell field
associated with a $U(1)$ subgroup of $E_{8}\times E_{8}$ 
or ${\it{Spin}(32)}/Z_{2}$.} defined by the 
metric
\begin{equation}
ds^2 = f(r) dt^2 - \frac{1}{h(r)} dr^2 -r^2d\Omega
\label{1.1}
\end{equation}
where,
\begin{eqnarray}
f(r) &=& \left( 1- \frac{2Me^{\phi_{0}}}{r}\right)
\left(1 - \frac{Q^{2}e^{3\phi_{0}}}{Mr}\right)^{-1}\nonumber\\
h(r) &=& \left(1-\frac{2Me^{\phi_{0}}}{r}\right)
\left(1- \frac{Q^{2}e^{3\phi_{0}}}{Mr}\right)
\label{2a}
\end{eqnarray}
with $\phi_{0}$ being the asymptotic 
constant value of the dilaton field.
We consider the case when $Q^{2}<2e^{-2\phi_{0}}M^{2}$
for which the above metric describes a blackhole with an event horizon 
situated at   
\begin{eqnarray}
r_{H}&=&2Me^{\phi_{0}}~.
\label{hor}
\end{eqnarray}
As mentioned earlier, with the aid of dimensional reduction technique,
the effective field theory  near the 
horizon becomes a two dimensional chiral theory 
with a metric given by the ``$r- t$" sector of the full spacetime metric
(\ref{1.1}) near the horizon.
The important point to note however is that the determinant of the GHS metric 
$\sqrt{(-g)} \neq 1$. 
The theory away from the horizon is not chiral and hence is anomaly free.

\noindent We now adopt the methodology in \cite{rbshailesh}.
For a two dimensional theory the expressions 
for the anomalous (chiral) and normal effective actions 
are known \cite{Isoumtwilczek, Leut}. 
We shall use only the anomalous form 
of the effective action for 
deriving the Hawking flux.
%\footnote{This is a slightly 
%different approach from the one
%adopted in \cite{rbshailesh} where they use only the anomalous (chiral)
%form of the effective action to determine the Hawking flux.}. 
Also we consider only the gravitational part of the effective action
since we have only gravitational anomaly in the region near the horizon
in the GHS case.
The energy-momentum tensor in the region near the horizon
is  computed by taking appropriate functional derivative of the chiral
effective action. Next, the parameters appearing
in the solution is fixed by imposing 
the vanishing of covariant energy-momentum
tensor at the horizon. Once these are fixed, 
the Hawking flux is  obtained by taking the asymptotic 
$(r\rightarrow {\infty})$ limit of the chiral energy-momentum 
tensor. Finally, we use the expression for the normal effective
action to establish a connection between the chiral and the
normal energy-momentum tensors.

\noindent With the above methodology in mind, 
we write down the gravitational part
of the anomalous (chiral) effective action (describing the theory 
near the horizon) \cite{Leut} 
\begin{equation}
\Gamma_{(H)}= -\frac{1}{3} z(\omega) 
\label{effaction}
\end{equation}
where $\omega_{\mu}$ is the spin connection and
\begin{equation}
z(v) = \frac{1}{4\pi}\int d^2x~ d^2y~ \epsilon^{\mu\nu}
\partial_\mu v_\nu(x) \Delta_{g}^{-1}(x, y)
\partial_\rho[(\epsilon^{\rho\sigma} + \sqrt{-g}g^{\rho\sigma})v_\sigma(y)]
\label{effaction1}
\end{equation}
where $\Delta_{g} = \nabla^{\mu}\nabla_{\mu}$ is the laplacian in this 
background.

\noindent The energy-momentum 
tensor is computed from a variation of this effective action.
To get their covariant forms in which we are interested, 
one needs to add appropriate local polynomials \cite{Leut}. 
Here we quote the final result for the covariant 
energy-momentum tensor \cite{Leut}: 
\begin{eqnarray}
{T^{\mu}}_{\nu} = \frac{1}{4\pi}\left(\frac{1}{48}D^{\mu}G D_{\nu}G 
-\frac{1}{24} D^{\mu} D_{\nu}G + \frac{1}{24}\delta^{\mu}_{\nu}R\right)
\label{9} 
\end{eqnarray}
where $D_{\mu}$ is the chiral covariant derivative
\begin{equation}
D_{\mu} = \nabla_{\mu} - \bar{\epsilon}_{\mu\nu}\nabla^{\nu}
= -\bar{\epsilon}_{\mu\nu}D^{\nu} \label{11}
\end{equation} 
and $\bar\epsilon^{\mu\nu}=\epsilon^{\mu\nu}/\sqrt{-g}$~, 
$\bar\epsilon_{\mu\nu}=\sqrt{-g}~\epsilon_{\mu\nu}$ are two
dimensional antisymmetric tensors for the upper and lower cases
with $\epsilon^{tr}=\epsilon_{rt}=1$.
Also $R$ is the two dimensional Ricci scalar given by 
\begin{eqnarray}
R=\frac{h~f^{''}}{f}+\frac{f^{'}h^{'}}{2f}
-\frac{f^{'2}h}{2f^2}
\label{1aa}
\end{eqnarray}
and $G$ is given by
\begin{eqnarray}
G(x) &=& \int d^2y  \ \Delta^{-1}_{g}(x,y)\sqrt{-g}~R(y)\label{5}
\end{eqnarray}
satisfying
\begin{equation}
\nabla^{\mu}\nabla_{\mu}G = R~.
\label{6}
\end{equation}
The solution for $G$ reads
\begin{equation}
G = G_o(r) - 4 pt + q \ ; \  \partial_{r}G_o = - \frac{1}{\sqrt{fh}}
\left(\sqrt{\frac{h}{f}}f'+z\right)
\label{8}
\end{equation}
where $p, q $ and $z$ are constants. 
Also note that $G_o$ is a function of $r$ only.

\noindent By taking the covariant divergence of (\ref{9}), 
we get the anomalous Ward identity
\begin{equation}
\nabla_{\mu}{T^{\mu}}_\nu = \frac{1}{96\pi} 
\bar{\epsilon}_{\nu\mu}\partial^{\mu}R~.
\label{13}
\end{equation}
The anomalous term is the covariant gravitational anomaly. This
Ward identity was also obtained from 
different considerations in \cite{shailesh}. 

\noindent In the region away from the horizon, the effective
theory is given by the standard effective action
$\Gamma$ of a conformal field
with a central charge $c=1$ 
in this blackhole background \cite{Isoumtwilczek} and reads:
\begin{eqnarray}
\Gamma =  \frac{1}{96\pi}\int d^2x d^2y \ \sqrt{-g}~ 
R(x)\frac{1}{\Delta_{g}}(x,y)\sqrt{-g}~R(y). 
\label{normaleff}
\end{eqnarray}
The energy-momentum tensor $T_{\mu\nu(o)}$ in the region outside 
the horizon is given by
\begin{eqnarray}
T_{\mu\nu(o)}&=&\frac{2}{\sqrt{-g}} \frac{\delta\Gamma}{\delta g^{\mu\nu}}
\nonumber\\
&=&\frac{1}{48\pi} \left(2g_{\mu\nu}R - 2\nabla_{\mu}\nabla_{\nu}G
+\nabla_{\mu}G\nabla_{\nu}G-\frac{1}{2}g_{\mu\nu}
\nabla^{\rho}G\nabla_{\rho}G\right)
\label{normalten}
\end{eqnarray}
and satisfies the normal Ward identity
\begin{equation}
\nabla_\mu {T^\mu}_{\nu(o)}=0~.
\label{normalward}
\end{equation}\\

%%%%%%%%%%%%%%%%%%%%%%%%%%%%%%%%%%%%%%%%%%%%%%%%%%%%%%%%%

%%%%%%%%%%%%%%% Energy Flux %%%%%%%%%%%%%%%%%%%%%%%%%%%%%
\noindent{\it Energy Flux:}\\

\noindent In this section we calculate the energy 
flux by using the expression for 
the covariant energy-momentum tensor (\ref{9}).
We will show that the results are the
same as that  obtained by the anomaly cancellation 
(consistent or covariant) method \cite{sun}. 
 
\noindent Using (\ref{11}) and the solution for $G(x)$ (\ref{5}), 
the $r-t$ component of the anomalous (chiral) covariant 
energy-momentum tensor (\ref{9}) becomes
\begin{eqnarray}
{T^{r}}_t(r)&=&\frac{1}{12\pi}\sqrt{\frac{h}{f}}
\left[p - \frac{1}{4}\left(\sqrt{\frac{h}{f}}f' + z\right)\right]^2 
\nonumber\\
&&+\frac{1}{24\pi}\sqrt{\frac{h}{f}}
\left[\sqrt{\frac{h}{f}}f'
\left(p-\frac{1}{4}\left(\sqrt{\frac{h}{f}}f' + z\right)\right) 
+ \frac{1}{4}hf'' - \frac{f'}{8}\left(\frac{h}{f}f'-h'\right)
\right].\label{18}
\end{eqnarray}
Now implementing the boundary condition 
namely the vanishing of the covariant energy-momentum tensor
at the horizon, ${T^{r}}_t(r_{H}) = 0$, leads to
\begin{equation}
p=\frac{1}{4}\left[z \pm \sqrt{f'(r_H)h'(r_H)}\right] \ ; 
\  f'(r_H) \equiv f'(r = r_{H})~. 
\label{19} 
\end{equation} 
Using either of the above solutions in (\ref{18}) yields
\begin{eqnarray}
{T^{r}}_t =\frac{1}{192\pi}\sqrt{\frac{h}{f}}
\left[f'(r_H)h'(r_H) - 
\frac{2h}{f}f'^{2} + 2hf''+f'h'\right]. 
\label{24}
\end{eqnarray} 
This expression is in agreement with that given in \cite{sun}. 

\noindent The energy flux is now given by the 
asymptotic ($r\rightarrow\infty$) limit of the
anomaly free energy-momentum tensor (\ref{normalten}). Now from 
(\ref{13}), we observe that the anomaly vanishes
in this limit. Hence the energy flux is abstracted
by taking the asymptotic limit of (\ref{24}).
This yields
\begin{equation}
{T^{r}}_t(r\rightarrow\infty)
=\frac{1}{192\pi}f'(r_H)h'(r_H)
\label{25}
\end{equation}
which correctly reproduces the Hawking flux \cite{das,sun}.

\noindent We now consider the normal (anomaly free) energy-momentum
tensor (\ref{normalten}) to establish its relation with
the chiral (anomalous) energy-momentum
tensor (\ref{24}). The $r-t$ component of ${T^{\mu}}_{\nu(o)}$
is given by
\begin{eqnarray}
{T^{r}}_{t(o)}(r)&=&-\frac{1}{12\pi}\sqrt{\frac{h}{f}}zp~.
\label{rtnormal}
\end{eqnarray}
The asymptotic form of the above equation (\ref{rtnormal}) must agree
with the asymptotic form of (\ref{18})\footnote{This is true
since the anomaly in the asymptotic limit ($r\rightarrow\infty$) 
vanishes as can be readily seen from (\ref{13}).}. This yields:
\begin{eqnarray}
p=-\frac{z}{4}~.
\label{asym}
\end{eqnarray}
Solving (\ref{19}) and (\ref{asym}) gives two solutions for $p$ and $z$:
\begin{eqnarray}
p&=&\frac{1}{8}\sqrt{f'(r_H)h'(r_H)}\quad;\quad 
z=-\frac{1}{2}\sqrt{f'(r_H)h'(r_H)}\nonumber\\
p&=&-\frac{1}{8}\sqrt{f'(r_H)h'(r_H)}\quad;\quad 
z=\frac{1}{2}\sqrt{f'(r_H)h'(r_H)}~. 
\label{solutions}
\end{eqnarray}
Using either of the above solutions in (\ref{18}) 
and (\ref{rtnormal}) yields (\ref{24}) and
\begin{eqnarray}
{T^{r}}_{t(o)}(r) =\frac{1}{192\pi}\sqrt{\frac{h}{f}}
f'(r_H)h'(r_H)~. 
\label{24aa}
\end{eqnarray} 
The above expressions (\ref{24}) and (\ref{24aa}) 
yields the equation between the chiral (anomalous) and the normal
energy-momentum tensors \cite{sun}.\\

\noindent{\it Discussions:}\\
In this paper, we have employed the effective action
approach (as was done in \cite{rbshailesh}) to derive
Hawking flux from the GHS blackhole in string theory.
As has been stressed in \cite{rbshailesh}, 
generally such approaches
require some other boundary condition 
apart from conditions at the horizon, as for example, 
the vanishing of ingoing modes at 
infinity \cite{Fulling, unruh}.
In this approach we only need covariant boundary conditions, the
importance of which was first stressed in \cite{rbshailesh}.
Another important ingredient in the entire analysis is the expression
for the anomalous (chiral) effective action 
(which yields anomalous Ward identity 
having covariant gravitational anomaly). 
The unknown parameters in the covariant energy-momentum tensor 
derived from this anomalous effective action 
were fixed by a boundary condition- namely the vanishing of the 
covariant energy-momentum tensor 
at the event horizon of the GHS blackhole.
Finally, the energy flux was extracted 
by taking the $r\rightarrow\infty$
limit of the chiral covariant energy-momentum tensor.
The relation between the chiral and the normal energy-momentum tensors
is also established by requiring that 
both of them match in the 
asymptotic limit which is possible since the anomaly
vanishes in this limit.\\
 
%%%%%%%%%%%%%%%%%%%%%%%%%%%%%%%%%%%%%%%%%%%%%%%%%%%%%%%%%%%%%%%%%%%%%%%%%%
\noindent {\it Acknowledgements:}\\
I thank Shailesh Kulkarni for useful discussions and the referee
for useful comments.  
%%%%%%%%%%%%%%%%%%%%%%%%%%%%%%%%%%%%%%%%%%%%%%%%%%%%%%%%%%%%%%%%%%%%%%%%%%

%%%%%%%%%%%%%%%%%%%%%%%%%%%%%%%%%%%%%%%%%%%%%%%%%%%%%%%%%%%%%%%%%%%%%%%%%%
\end{document}